\documentstyle[11pt,aaspp4]{article}

\def\ltsima{$\;\buildrel < \over \sim \;$}
\def\simlt{\lower.5ex \hbox{\ltsima}}
\def\lesssim{\simlt}
\def\gtsima{$\;\buildrel > \over \sim \;$}
\def\simgt{\lower.5ex \hbox{\gtsima}}
\def\gtrsim{\simgt}
\begin{document}
\title{The Ratio of Ortho- to Para-H$_2$ in Photodissociation Regions}

\author{Amiel Sternberg$^{1,2}$ and David A. Neufeld$^3$}

\vskip 0.5 true in {\parskip 6pt

\noindent{$^1$ School of Physics and Astronomy,
Tel Aviv University, Ramat Aviv 69978, Israel}

\noindent{$^2$ Department of Astronomy, 
University of California, Berkeley, CA, 94720-3411}
  
\noindent{$^3$ Department of Physics \& Astronomy,  
The Johns Hopkins University, 3400 North Charles Street,  
Baltimore, MD 21218}}
   
\vskip 1 true in


\begin{abstract}

We discuss the ratio of ortho- to para-H$_2$ in photodissociation
regions (PDRs). We draw attention to an apparent confusion in the
literature between the ortho-to-para ratio of molecules
in FUV-pumped vibrationally {\it excited}
states, and the {\it total} H$_2$ ortho-to-para abundance ratio.
These ratios are not the same 
because the process of FUV-pumping 
of fluorescent H$_2$ emission in PDRs
occurs via {\it optically thick} absorption lines.
Thus, gas with an equilibrium ratio of ortho- to para-H$_2$ 
equal to 3 will yield FUV-pumped vibrationally
excited ortho-to-para ratios
smaller than 3, because the ortho-H$_2$ pumping rates are
preferentially reduced by optical depth effects.
Indeed, if the ortho and para pumping lines are on the
``square root'' part of the curve-of-growth, 
then the expected ratio of ortho and para vibrational
line strengths is $3^{1/2} \sim 1.7$,
close to the typically observed value.
Thus, contrary to what has sometimes been stated in the 
literature, most previous measurements of the ratio of ortho- to 
para-H$_2$ in vibrationally excited states are entirely
consistent with a total ortho-to-para ratio of 3, the equilibrium
value for temperatures greater than 200 K.
We present an analysis and several detailed models which illustrate 
the relationship
between the total ratios of ortho- to para-H$_2$ and the 
vibrationally excited ortho-to-para ratios in PDRs.
Recent {\it Infrared Space Observatory (ISO)} measurements
of pure rotational and vibrational
H$_2$ emissions
from the PDR in the star-forming region S140 provide
strong observational support for our conclusions.

\end{abstract}

\section{Introduction}

Infrared (IR) vibrational and rotational 
emission lines of molecular hydrogen (H$_2$) have
been observed in a wide range of interstellar environments,
including molecular clouds in star-forming regions,
Herbig-Haro objects, reflection nebulae, and planetary nebulae.
H$_2$ emission lines have also been observed in many starburst galaxies,
and active galactic nuclei.  Two principal sources of such
emissions are (1) photodissociation regions (PDRs) where the molecules
are vibrationally excited by far-ultraviolet (FUV) pumping
or collisionally excited in gas heated by FUV radiation;
and (2) shocked regions, in which the hydrogen molecules are
collisionally excited in hot gas behind the shock waves.
Observations of the rich H$_2$ vibrational spectrum yield line
ratios that are valuable probes of the physical conditions within the
emitting source: the relative strengths of the $v$=1--0 and $v$=2--1
bands, for example, allow fluorescent excitation produced by
FUV-pumping to be distinguished
from collisional excitation (Black \& Dalgarno 1976;
Black \& van Dishoeck 1987;
Sternberg 1988; Sternberg \& Dalgarno 1989;
Burton, Hollenbach \& Tielens 1990; Draine \& Bertoldi 1996). 

The relative 
strengths of emissions from ortho- and para-H$_2$ molecules have
been the subject of intensive observational investigation.
In outflow regions where shock excitation is the primary
emission mechanism, observations of vibrational emission lines
typically reveal ortho-to-para ratios for
vibrationally-excited states that are comparable to 3 
(Smith, Davis \& Lioure 1997), as expected if the gas behind such shocks
has a ratio of  ortho- to para-H$_2$ in local thermodynamic
equilibrium (LTE) at high temperature ($T \simgt 200$~K).
\footnote{Note, however, that recent observations of
{\it pure rotational} emissions from the outflow source HH 54
(Neufeld, Melnick \& Harwit 1998) imply a non-equilibrium
ortho-to-para H$_2$ abundance ratio $\sim 1.2$ in gas of temperature
of 650~K.  The observed ratio of ortho- to para-H$_2$
in HH 54 is believed to be the remnant of an
earlier phase in the thermal history of the gas when
the temperature was $\simlt 90$~K.
As far as we are aware, HH 54 is the {\it only} astronomical source
observed to date in which the total ratio of ortho- to para-H$_2$ is
demonstrably different from the expected LTE value.}
PDRs, by contrast, often exhibit ortho-to-para ratios for
vibrationally excited states
that are smaller than 3, with typical values in the range 1.5 -- 2.2
(Hasegawa et al.\ 1987,
Ramsay et al.\ 1993, Chrysostomou et al.\ 1993, Hora \& Latter 1996;
Shupe et al. 1998).

In this paper we re-examine the interpretation of the 
ortho-to-para ratios for H$_2$ in PDRs.
Previous discussions have
failed to make a critical distinction between
(1) the ortho-to-para ratio for vibrationally excited states
populated by FUV-pumping;
and (2) the true ortho-to-para abundance ratio, which is
generally dominated by molecules in the ground vibrational state.
In \S 2 we present a brief review of H$_2$ excitation
in PDRs. We then argue that the ortho-to-para ratio for vibrationally excited
H$_2$ molecules in PDRs is generally not equal
to the true ortho-to-para ratio,
because the process of FUV-pumping occurs via {\it optically thick}
absorption lines and the ortho pumping rates are therefore
preferentially reduced by larger transition optical depths.
In \S 3 we present several detailed model computations
which illustrate the behavior of the ortho-to-para 
ratios in PDRs for a range of conditions.
In \S 4 we analyze recent {\it Infrared Space Observatory (ISO)}
observations of the PDR in the star-forming
region S140 which  provide strong support for our conclusions.
In \S 5 we present a discussion and summary.

\section{Theory}

\subsection{H$_2$ in PDRs}

PDRs are produced at the boundaries of molecular clouds which are 
exposed to FUV radiation fields ($\sim 6-13.6$ eV).
The incident radiation heats the gas, and drives the cloud
chemistry via photodissociation of molecules and photoionization
of trace species. Recent reviews of observational and theoretical
studies of PDRs have been presented by Hollenbach \& Tielens (1997, 1998)
and Sternberg (1998). 

Photodissociation and FUV-pumping of hydrogen molecules are
critical processes in PDRs.  At the cloud edges the H$_2$
molecules are photodissociated rapidly, and the hydrogen gas
is mainly atomic.  The FUV radiation is attenuated with increasing
cloud depth, and eventually photodissociation becomes slow
compared to grain-surface H$_2$ formation, and the hydrogen
becomes molecular.  In equilibrium clouds, molecular destruction
is balanced by formation at every location, and
$$
Dn({\rm H}_2) = Rnn({\rm H})                             \eqno(1)
$$
where $n({\rm H})$ and $n({\rm H}_2)$ are the
local atomic and molecular hydrogen volume densities (cm$^{-3}$),
$n=n({\rm H})+2n({\rm H}_2)$ is the total density of hydrogen nuclei,
$R$ is the grain-surface H$_2$ formation rate coefficient (cm$^3$ s$^{-1}$)
and $D$ is the local H$_2$ photodissociation rate (s$^{-1}$).
At each point in the cloud the populations of H$_2$ molecules 
in specific vibrational and rotational levels (identified
by the quantum numbers $v$ and $j$) are determined by the competing
processes of
FUV-pumping, molecular formation, collisional excitation and
de-excitation, and radiative decay.

H$_2$ photodissociation and FUV-pumping
occur via the absorption of Lyman and Werner band photons
in electronic transitions from the $X^1\Sigma_g^+$ ground state
to the excited $B ^1\Sigma_u^+$, $C^1\Pi_u^+$, and $C^1\Pi_u^-$
states.  These excitations are followed
by rapid spontaneous radiative transitions to the continuum and to 
bound $vj$ levels of the ground electronic state.
The transitions to the continuum lead to dissociation.
The FUV-pumped molecules
decay in a cascade of radiative quadrupole vibrational transitions
which produces the characteristic spectrum
of near-red and infrared fluorescent emission lines
observed in PDRs (Black \& Dalgarno 1976;
Black \& van Dishoeck 1987; Sternberg 1988; Neufeld \& Spaans 1996;
Draine \& Bertoldi 1996).

The Lyman and Werner band absorption lines are expected to
become optically thick in PDRs, and the molecules are
then said to ``self-shield".  The typical absorption line 
becomes optically thick at an H$_2$ column density
of $\sim 10^{14}$ cm$^{-2}$, which is much smaller than the
hydrogen column of $\sim 5\times 10^{20}$ cm$^{-2}$ at
which FUV dust-opacity becomes significant for normal interstellar
gas-to-dust ratios.  
Indeed, 
because the strongest FUV absorption lines become saturated at
H$_2$ columns $\gtrsim 10^{17}$ cm$^{-2}$, these lines 
lie on the square-root part of the curve-of-growth
throughout most of the PDR and FUV pumping and photodissociation
are initiated by absorptions in the ``damping wings" of the lines. 
At sufficiently large cloud depths
the FUV-pumping and photodissociation rates become negligibly 
small due to the combined  effects of dust-opacity 
and overlap of the broadest absorption lines
(Tielens \& Hollenbach 1985; 
Sternberg 1988; Draine \& Bertoldi 1996).

Collisions play a critical role in the 
excitation of the lowest lying $j$-levels of the ground ($v=0$)
vibrational state.
The most important are inelastic collisions with other
H$_2$ molecules,
hydrogen atoms, and protons. 
In high-density clouds the $vj$ level populations of
vibrationally excited molecules are influenced by collisional
processes $\rm H+H_2(vj)\rightarrow H+H_2(v^\prime j^\prime)$ and 
$\rm H_2+H_2(vj)\rightarrow H_2 + H_2(v^\prime j^\prime)$
which induce vibrational energy transfers.
The relative intensities of the resulting H$_2$ emission
lines differ in high density
clouds because such collisional processes
become more effective
relative to
radiative decay
with increasing cloud density
(Sternberg \& Dalgarno 1989; Burton, Hollenbach \& Tielens 1990;
Draine \& Bertoldi 1996).

Molecular formation can also influence the populations in excited $vj$ states. 
Although in 
equilibrium PDRs only one molecular formation event occurs 
for every $\sim 9$ FUV-pumping excitations, formation can dominate
the excitation of molecules in specific $vj$ levels that are
not directly populated by FUV-pumping. 

\subsection{Ortho-to-Para Ratio}

The ratio of ortho- to para-H$_2$ is a critical parameter in PDRs.
We recall that
ortho-H$_2$ possesses a total nuclear spin of 1 and
exists only in states of odd rotational quantum number,
while para-H$_2$ has a zero total nuclear spin and
is represented only by states of even $j$. 
Ortho- and para-H$_2$ are not interconverted 
by any radiative process.  In particular, the
Lyman and Werner band transitions and the quadrupole
ro-vibrational transitions which occur in PDRs 
cannot induce ortho-para conversions.
Following the formation of H$_2$ in each of the ortho and
para modifications, conversions can occur in the gas
via ``spin exchange" collisions. In cold (T$\lesssim 100$ K) gas
the ortho-para conversion occurs slowly via collisions with protons
(Dalgarno, Black \& Weisheit 1973; Flower \& Watt 1987)
$$
\rm H_2(ortho) + H^+ \rightleftharpoons  H_2(para) + H^+  \ \ \ ,
$$
where in PDRs a small proton abundance is maintained by cosmic-ray
or X-ray ionization (Sternberg \& Dalgarno 1995; Maloney, Hollenbach
\& Tielens 1996). In warmer and partially dissociated
gas, reactive collisions with
hydrogen atoms (Martin \& Mandy 1993; Tin\'e et al. 1997;
Lepp et al. 1998)
$$
\rm H_2(ortho) + H \rightleftharpoons H_2(para) + H   \ \ \ 
$$
control the ortho-para conversion. 
Grain surface exchange reactions are a third possible mechanism
for ortho-para conversion: we exclude such reactions from 
further consideration, because their efficiency is quite uncertain
and because they are unlikely to affect significantly the ortho to para-
ratio under conditions where vibrational emissions are excited
(Burton, Hollenbach \& Tielens 1992).

When they are sufficiently rapid the
ortho-para converting collisions with H or H$^+$
drive the ortho-to-para abundance ratio
to local thermodynamic equilibrium.
In Fig.\ 1 we plot the LTE ortho-to-para ratio, $\alpha(T)$, as a function
of temperature (see also Figure 4 of Burton et al.\ 1992). For $T\gtrsim 200$ K, 
$\alpha =  3$.  At lower temperatures
the molecules are driven into the $j=0$ (para) ground state,
and $\alpha$ falls below 3. 

The actual ortho-to-para H$_2$ abundance ratio in PDRs depends on the
competing effects of collisions, selective photodissociation
of ortho and para molecules, and the molecular formation
process.  Furthermore, as we will now argue, the ortho-to-para
ratio of molecules in vibrationally excited states is generally {\it not}
equal to the true ortho-to-para H$_2$ abundance ratio.

The behavior may be understood by considering a simplified
analytic model in which FUV-pumping and photodissociation
is assumed to occur via
a single pair of (non-overlapping) ortho and para absorption lines.
The ortho-to-para ratio is controlled by 
the microscopic formation and destruction processes for the ortho- and
para-H$_2$.
In equilibrium
$$
D_on_o + q(o\rightarrow p)n_xn_o = R_onn({\rm H}) + q(o\leftarrow p)n_xn_p   \eqno(2)
$$
and
$$ 
D_pn_p + q(o\leftarrow p)n_xn_p = R_pnn({\rm H}) + q(o\rightarrow p)n_xn_o   \eqno(3)
$$
where $n_o$ and $n_p$ are the local ortho- and para-H$_2$ volume densities
(cm$^{-3}$), $D_o$ and $D_p$ are the local dissociation rates (s$^{-1}$)
of the ortho and para molecules, 
$q(o\rightleftharpoons p)$ are the rate coefficients (cm$^3$ s$^{-1}$)
for ortho-para conversions induced by
collisions with partners with density $n_x$,
and $R_o$ and $R_p$ are the ortho and para 
molecular formation rate
coefficients.  The total hydrogen density $n=n({\rm }H)+2n({\rm H}_2)$ where
$n({\rm H}_2)=n_o+n_p$.  Equation (1) for the $\rm H/H_2$ abundance ratio 
is the sum of expressions (2) and (3), where
$Dn({\rm H}_2)=D_on_o+D_pn_p$, and $R=R_o+R_p$.
Because most of the hydrogen molecules are almost always
in the ground vibrational state and only a very small fraction
are in vibrationally excited states,
$D_o$ and $D_p$ are dissociation rates out of the $v=0$ level.

If ortho-para converting
collisions in the $v=0$ level 
are very rapid compared with either molecular formation
or photodissociation then it follows from eqns. (2) and (3), 
and the principle of detailed balance, that
the {\it total} ortho-to-para abundance ratio
$$
{n_o \over n_p}={q(o\leftarrow p) \over q(o\rightarrow p)} 
  = \alpha(T_{gas})\eqno(4)
$$
and the ortho-to-para ratio is driven to LTE.  The detailed numerical 
calculations presented in \S 3 and \S 4 below suggest that the case of rapid
ortho-para conversion amongst the $v=0$ states does indeed apply 
throughout most of the PDR under most conditions of astrophysical 
interest.  It is nevertheless instructive to consider the opposite
limit in which ortho-para conversion is negligibly slow.

If ortho-para converting collisions are slow ($q=0$) then
$$
{n_o \over n_p}={R_o \over R_p}{D_p\over D_o}
=  {D_p \over D_o}\alpha(T_{form}) \ \ \  \eqno(5)
$$
where $T_{form}$ is the formation temperature.
\footnote{
Here we make the simplifying (though not necessary) assumption that
the grain surface formation process produces the H$_2$ in an LTE
population distribution characterized by a formation temperature $T_{form}$,
leading to an ortho-to-para ratio $\alpha(T_{form})$.}
When the FUV absorption lines 
are optically thin, then $D_p/D_o = 1$ and the total ortho-to-para ratio
equals $\alpha(T_{form})$. However, as the absorption lines become optically
thick, $D_p/D_o$ can become very
large, or very small, depending on which set of ortho or para
absorption lines first become optically thick.
For absorption lines on the damping wings
$D_p/D_o=(N_p/N_o)^{-1/2}$, where $N_o$ and $N_p$ are
the total column densities of ortho and para molecules,
and equation (5) can be written as the differential equation
$$
{dN_o \over dN_p}\equiv{n_o \over n_p}=\alpha(T_{form})
\Bigl({N_o \over N_p}\Bigr)^{1/2} \ \ \ .           \eqno(6)
$$
If $\alpha(T_{form})$ is constant through the PDR it follows that 
$$
{n_o \over n_p}={N_o \over N_p}=\alpha^2(T_{form})  \ \ \ . \eqno(7)
$$
Thus, when ortho-para converting collisions are ineffective,
preferential shielding of the ortho or para molecules maintains
an ortho-to-para abundance ratio equal to the {\it square} of the 
value set by molecular formation.

We now consider the ortho-to-para ratio for vibrationally
excited states.  In the spirit of the two-line approximation used to treat 
the FUV-pumping and photodissociation in our simplified analytic
model, we lump all possible vibrationally-excited states into
just two states, one representing vibrationally-excited ortho-H$_2$
and one representing vibrationally-excited para-H$_2$.  

If ortho-para
conversion {\it amongst vibrationally-excited states} were rapid
relative to other relevant processes (as is usually the case for
ortho-para conversion for $v=0$ molecules), then the
ortho-to-para ratio in vibrationally-excited states would 
also equal the LTE value, $\alpha (T_{gas})$.  However, 
when FUV-pumping dominates the molecular excitation, ortho-para
conversion amongst vibrationally-excited states is generally slow, leading
to more interesting behavior.

We assume that
the vibrationally excited ortho-
and para-H$_2$ molecules are populated by FUV pumping,
and are removed by quadrupole decay.
This is the limit of ``radiative fluorescent H$_2$ emission'' 
(Sternberg \& Dalgarno 1989), which applies 
when (1) collisional excitation is negligible relative to FUV-pumping;
and (2) when collisional de-excitation and photodissociation are
negligible removal mechanisms relative to spontaneous radiative decay. 

In equilibrium
$$
n_o^* = P_on_o/A     \eqno(8)
$$
and
$$
n_p^* = P_pn_p/A   \eqno(9)
$$
where $n_o^*$ and $n_p^*$ are the volume densities of vibrationally
excited ortho- and para-H$_2$, $P_o$ and $P_p$ are the ortho and
para FUV-pumping rates, 
and $A$ is the quadrupole radiative
decay rate (which does not depend on the ortho or para 
character of the molecule).
The sum of these expressions
is the formation-destruction equation for vibrationally excited molecules
$$
n^* = Pn({\rm H}_2)/A       \eqno(10)
$$
where $n^*$ is the total density of vibrationally excited molecules,
and $P$ is the total (net) FUV-pumping rate.

It follows from (8) and (9) that
$$
{n_o^* \over n_p^*}={P_o\over P_p}{n_o\over n_p}       \ \ \ . \eqno(11)
$$
For absorption lines on the damping wings, $P_o/P_p=(N_o/N_p)^{-1/2}$, so that
$$
{n_o^* \over n_p^*}=
{n_o \over n_p}\Bigl({N_o\over N_p}\Bigr)^{-1/2}         \eqno(12)
$$
If the true ortho-to-para ratio $n_o/n_p$ is in LTE and is equal to a fixed
value $\alpha(T_{gas})$, then $N_o/N_p=n_o/n_p$, and
$$
{N_o^* \over N_p^*}=\Bigl({N_o \over N_p}\Bigr)^{1/2}
 = \sqrt{\alpha(T_{gas})}   \ \ \ .   \eqno(13)
$$
If collisional processes are unable
to maintain the true ortho-to-para ratio in LTE,
it follows from eqns. (5) and (11), and the fact that
$P_o/D_o=P_p/D_p$ are constant molecular branching ratios,
\footnote{We note that the Lyman and Werner band transition oscillator
strengths do not depend on the ortho or para
character of the H$_2$ molecule.}
that the vibrationally excited ortho-to-para ratio
$$
{n_o^* \over n_p^*}=\alpha(T_{form})     \ \ \ .\eqno(14)
$$
It then follows from eqn. (12) that
$$
{N_o\over N_p}=\Bigl({N_o^* \over N_p^*}\Bigr)^2=\alpha^2(T_{form})
     \ \ \   \eqno(15)
$$
assuming $\alpha(T_{form})$ is constant through the PDR.

Equations (13) and (15) represent the key result of our paper.
They show that when collisional processes are ineffective
at thermalizing the ortho-to-para ratio
in vibrationally {\it excited} states, then in the 
limit of radiative fluorescent H$_2$ emission
this ratio will roughly equal the {\it square root} of the
true ortho-to-para abundance ratio. In particular, if the ($v=0$) 
ortho-to-para abundance ratio 
is everywhere equal to an LTE value of 3, the vibrationally 
excited ortho-to-para ratio
will tend to a non-LTE value of $\sqrt{3}\approx 1.7$, because of the lower 
rates of FUV-pumping in the optically thicker ortho absorption lines.
Alternatively, if ortho-para interconversion of the $v=0$ molecules
is inefficient then the vibrationally excited ortho-to-para
ratio is fixed by the molecular formation process. If
the ortho-to-para ratio set by formation is equal to 3, the
ortho-to-para abundance ratio will then tend 
at large cloud depths
to a non-LTE ratio of 9, due
to selective self-shielding of the ortho H$_2$.

Equations (13) or (15) describe the relationship between the
vibrationally excited and total ortho-to-para ratios 
for the case of radiative fluorescent H$_2$ emission where
FUV-pumping dominates the vibrational excitation and
the excited molecules are removed by radiative decay.
%
%
However, in dense ($n\gtrsim 5\times 10^4$ cm$^{-3}$) clouds 
exposed to intense FUV-fields ($\gtrsim 10^4$ times the intensity
of the mean interstellar field)
the PDRs can become hot ($T_{gas} \gtrsim 1000$ K), and collisional excitation
rather than FUV-pumping then dominates the  
vibrational excitation, particularly of $v=1$ molecules
(Sternberg \& Dalgarno 1989; Burton et al. 1990).
Under these conditions the vibrationally excited ortho-to-para ratio
will equal the true ortho-to-para ratio since the optical depth
effects associated with FUV-pumping do not apply. 
Ortho-para conversion
is expected to be rapid in hot PDRs so the vibrationally
excited ortho-to-para ratio for the collisionally excited
molecules will tend to the LTE value of 3.
The observations indicate that most PDRs are generally
either in the limit of radiative fluorescent H$_2$ emission
for which equations (13) and (15) apply, or are those which
are hot enough to produce collisionally excited vibrational
H$_2$ emission. Nevertheless two additional limits may be considered.
First, in sufficiently dense gas, the cascade of vibrational
de-excitations which follows FUV-pumping can become dominated by
collisional rather than radiative de-excitations (even when
collisional excitation is slow). In this
limit of ``collisional fluorescent H$_2$ emission" de-excitations
via H-H$_2$ collisions can lead to ortho-para 
interconversion of the vibrationally excited molecules.
Second, if the FUV-field is sufficiently intense that
photodissociation dominates the removal of the vibrationally
{\it excited} molecules, the distinction between the vibrationally
excited and $v=0$ ortho-to-para ratios disappears, 
and the vibrationally excited ortho-to-para
ratio is then also affected by the preferential self-shielding of
vibrationally-excited ortho-H$_2$ molecules. 
The ortho-to-para ratio in any vibrational state (whether the
ground state or an excited state) is 
then simply given by equation (4)
if ortho-para conversion is rapid, or equation (7)
if ortho-para conversion is slow. 

We now present detailed model computations which
illustrate the behavior of the ortho-to-para ratios in the 
limits described by equations (13) and (15).

\section{Models}

We have carried out a series of computations using an
updated version of the PDR models described by
Sternberg \& Dalgarno (1989, 1995).  

Our models consist
of static, plane-parallel, semi-infinite clouds which
are exposed, on one side, to isotropic FUV radiation fields.
We adopt FUV fields with spectral shapes identical to the Draine's (1978) 
fit to the mean interstellar field, multiplied by an
intensity scaling factor $\chi$.
At each cloud depth $z$ (cm) we solve
equation (1) for the $\rm H/H_2$ ratio, and compute the steady-state
population densities in each of the 301 rotational and
vibrational H$_2$ levels in the ground electronic state.
In our models we adopt
the level energies and quadrupole transition rates
given by Wolniewicz, Simbotin \& Dalgarno (1998),
and we employ the Lyman and Werner band
oscillator strengths and dissociation probabilities
listed by Allison \& Dalgarno (1970), and Stephens \& Dalgarno (1972).
We treat the opacity and self-shielding in each of the several
hundred Lyman and Werner absorption lines using the analytic formulae 
provided by Federman, Glassgold \& Kwan (1979).
We assume a frequency
independent FUV continuum
dust
extinction
cross section per hydrogen nucleus $\sigma = 2\times 10^{-21}$ cm$^2$
(Sternberg \& Dalgarno 1995).
As discussed by Tielens \& Hollenbach (1985) and
by Draine \& Bertoldi (1996), absorption line overlap begins to
significantly affect the self-shielding 
when the H$_2$ column density exceeds
$\sim 3\times 10^{20}$ cm$^{-2}$.  However, at such
large column densities dust opacity becomes large ($\tau \gtrsim 0.6$ for
our choice of $\sigma$) and the FUV-pumping rates are in any case
reduced exponentially. We therefore 
ignore the effects of line overlap in the models 
presented here.

In our models we assume
that the grain surface molecular formation rate coefficient
$R=3\times 10^{-18}T_{gas}^{1/2}$. We also assume
that one third of the H$_2$ binding energy is released as
internal rotational and vibrational excitation, in an initial LTE
$vj$ level distribution corresponding to a formation temperature
$T_{form}=(1/3)E_{diss}/k = 1.73\times 10^4$ K, where $E_{diss}$ 
is the dissociation
energy for H$_2$ 
(Black \& Dalgarno 1976; Black \& van Dishoeck 1987; 
Sternberg \& Dalgarno 1989).
For this choice,
$\alpha(T_{form})=3$.

In Figs.\ 2--4 we display the results of three illustrative model computations.
For each model
we plot the atomic hydrogen fractions, $n(H)/n$, and the
column densities of FUV-pumped molecules, $N^*/N^*_{max}$, as functions of the
visual extinction $A_V$ measured from the cloud surface.
In each model the gas is fully atomic at the cloud edge, and at
large cloud depths the atomic fractions become small. 
The column densities
of vibrationally excited molecules, $N^*$, increase linearly with
depth through the atomic layer (where the FUV-pumping and H$_2$ formation
rates per unit volume are constant) and then approach asymptotic
values, $N^*_{max}$, at large cloud depths where the pumping rates become
vanishingly small.  
For each model we also plot, as functions of $A_V$, 
the local values of the total ortho-to-para abundance ratio ($n_o/n_p$);
the ratios of the total column densities
in ortho and para molecules ($N_o/N_p$);
the local ortho-to-para ratio for vibrationally
excited molecules ($n^*_o/n^*_p$); and the ratios of the column
densities in vibrationally excited ortho and para molecules ($N^*_o/N^*_p$).
For each model we also display ``excitation diagrams" in which
we plot the values of ${\rm log}_{10}(N_{vj}/g)$ vs. $E_{vj}$,
where $N_{vj}$ are the total cloud column densities in each $vj$ level,
and $E_{vj}$ and $g$ are the level energies and
statistical weights.  For ortho-H$_2$ $g=3(2j+1)$, and
for para-H$_2$ $g=2j+1$.  For clarity 
only the lowest $j=0-7$ rotational levels in each vibrational state
are displayed in the excitation diagrams.

In all three models we adopt a
total hydrogen particle density $n=10^4$ cm$^{-3}$, an
FUV field intensity $\chi=2\times 10^3$, and assume a uniform gas temperature
$T_{gas}=500$ K. The values of $n$ and $\chi$ are representative
of typical PDRs (Hollenbach \& Tielens 1998). The gas temperature of 500 K is suggested 
by recent observations of pure rotational lines of H$_2$ in several
sources including the PDR in S140 (Timmermann et al.\ 1996; see \S 4).
All three of these models are in the limit of radiative fluorescent emission discussed
in \S 2.2.

In our first model we force the rotational
level populations in the ground ($v=0$) vibrational level to LTE
everywhere in the cloud. This is equivalent
to assuming that collisional ortho-para conversion is everywhere
much more rapid than photodissociation or molecular
formation for the $v=0$ molecules. 
With this assumption the ortho-to-para ratio in the
$v=0$ level is equal to exactly 3 everywhere in the cloud.
In this model we make the further assumption that ortho-para interconversion
does {\it not} occur for vibrationally excited molecules.
That is, we assume that ortho-para converting collisions are slow
compared to all of the other processes which populate or depopulate
the vibrationally excited molecules.
The results are shown
in Fig. 2 which shows that the total ortho-to-para ratio
(both $n_o/n_p$ and $N_o/N_p$)
is, in fact, everywhere very close to 3. This simply reflects the
fact that the total ortho-to-para ratio is dominated by
$v=0$ molecules for which the ortho-to-para ratio is fixed at 3
by assumption. 
However, the vibrationally
excited ortho-to-para ratio falls below 3, and at large depths approaches an
asymptotic value of $\sim 1.7$.  The difference between the total and vibrationally
excited ortho-to-para ratio is also illustrated by the excitation diagram
for this model. In this model the values of
${\rm log}(N_{0j}/g)$ lie on a single straight line (with slope
proportional to $1/T_{gas}$)
as expected for a rotational level distribution in LTE.
On the other hand, the ($j=0-7$) rotational distributions for 
vibrationally excited ortho and para molecules are each 
characterized by rotational temperatures $T_{rot}\sim 1500$ K,
as is typical for fluorescent emission (e.g. Sternberg \& Dalgarno 1989;
Draine \& Bertoldi 1996). However, the distinct
ortho and para distributions in the excitation diagram are vertically 
displaced from one another by a factor
$\sim 1.7$ which corresponds to the (low) non-LTE
ortho-to-para ratio for the vibrationally excited molecules. 
This model demonstrates explicitly that an ortho-to-para ratio of
$\sim 1.7$ in vibrationally excited states is consistent with 
a true LTE ortho-to-para abundance ratio of 3. This behavior
is consistent with equation (13).

Next, we consider a model in which ortho-para interconversions
do not occur even for $v=0$ molecules. The results are shown in Fig. 3.  
The vibrationally {\it excited} ortho-to-para ratio is now
everywhere very close to a value of 3, whereas the total ortho-to-para
ratio (dominated by the $v=0$ molecules) varies markedly through
the cloud.
In this model the vibrationally excited ortho-to-para ratio is set
by molecular formation only (with $\alpha(T_{form})=3$).
However, the ortho-to-para ratio for $v=0$ molecules is also significantly
affected by preferential self-shielding of the ortho molecules.
Thus, at the cloud edge, where the ortho and para FUV absorption
lines are all optically thin, the total ortho-to-para ratio equals the
formation value of 3. At a certain cloud depth the ($v=0$) ortho molecules
begin to self-shield, while the para molecules continue to be
dissociated rapidly, and the
total ortho-to-para ratio becomes large.
In our model the ortho-to-para abundance ratio
reaches a maximum value of $\sim 500$ at
$A_V\sim 0.1$.  At still larger cloud depths the dominant ortho
and para absorption lines become saturated, and the total ortho-to-para
abundance ratio approaches a value $\sim 3^2=9$. 
The excitation diagram for this model 
shows that the populations distributions for
vibrationally excited ortho and para molecules now lie on
single straight lines consistent
with an ortho-to-para ratio of 3. However, the
rotational level distribution for $v=0$ molecules now
display a non-LTE pattern, 
corresponding to the integrated ortho-to-para abundance ratio of $\sim 9$.
The behavior illustrated by this model is 
consistent with equation (15).

In the third model of this series we consider a more realistic
computation in which we explicitly include ortho-para
conversions via $\rm H^+-H_2$ and $\rm H-H_2$ collisions,
for both $v=0$ and vibrationally excited molecules.
In this model we adopted the rate coefficients computed by Gerlich (1990) 
for $\rm H^+-H_2$ collisions (see also Tin\'e et al. 1997),
and we assumed a
proton density $n({\rm H}^+)=10^{-5}n$ in the atomic zone,
decreasing to $10^{-7}n$ in the molecular zone.
For the $\rm H-H_2$ collisions
we adopted the semi-classical rate coefficients computed by Lepp et al. (1998).
\footnote{The Lepp et al. (1998) rate-coefficients were computed
using the BKMP potential (Boothroyd et al. 1996).  Martin \& Mandy (1993)
presented similar computations using the less accurate LSTH potential
(Truhlar \& Horowitz 1978).} 
The results are shown in Fig. 4. In this model the ortho-to-para
conversion rates vary from $\sim 10^{-13}n=10^{-9}$ s$^{-1}$ 
due to $\rm H-H_2$ collisions
in the atomic zone,
to $\sim 10^{-10}n({\rm H}^+)=10^{-13}$ s$^{-1}$ 
due to $\rm H^+-H_2$ collisions
in the molecular zone.
These rates are much smaller
than the unshielded photodissociation rate
$\sim 5\times 10^{-11}\chi \rm\, s^{-1}=10^{-7}$ s$^{-1}$ at the cloud surface.
Thus, an intermediate cloud
layer (near $A_V=0.1$) exists in which the total ortho-to-para ratio becomes large
due to preferential shielding of the ortho molecules. At still 
larger cloud depths the photodissociation rates are attenuated
to values smaller than the ortho-para conversion rate
and the local ortho-para abundance
ratio is thermalized to a value of 3. 
The vibrationally excited molecules, on the other hand, are nowhere
able to attain a thermal ratio of 3, 
because the radiative decay rates ($\sim 10^{-6}$ s$^{-1}$)
are much faster than the ortho-para conversion rates.
Thus, the vibrationally excited ortho-to-para ratio is set mainly
by the relative rates of FUV pumping of the ortho and para molecules,
and the local vibrationally excited
ortho-to-para ratio, $n_o^*/n_p^*$, tends to a value of $3^{1/2}\approx 1.7$
due to the optically thicker ortho absorption lines.
The ratio of vibrationally excited 
ortho and para column densities,
$N_o^*/N_p^*$, converges to the slightly larger value of 2.2,
due to the small (but non-negligible) 
effects of collisional
ortho-para interconversions of vibrationally excited molecules 
in the atomic hydrogen zone. 
The excitation diagram for this model is very similar
to the excitation diagram for the ``$v=0$ in LTE" model
shown Fig. 2. In both models, the rotational populations for
the vibrationally excited ortho and para molecules form
distinct and vertically displaced distributions, consistent
with the low and non-LTE vibrationally excited ortho-to-para ratios.
The excitation diagram in Fig. 4 shows that
the column densities in the lowest lying
($j\le 4$) rotational
levels of the $v=0$ molecules are very close to LTE, but that 
when realistic collisional rate coefficients are assumed the
higher lying rotational levels become progressively
subthermally populated.   
This model illustrates explicitly
that ortho-para conversion may typically be rapid enough
to maintain an LTE ortho-to-para abundance ratio of 3 for the
$v=0$ molecules, but not for vibrationally excited molecules.
The preferentially lower ortho pumping rates therefore drive the
vibrationally excited ortho-to-para ratio to the non-LTE value 
of $\sqrt{3}$, consistent with equation (13).

\section{Observations}

Vibrational emissions from H$_2$ molecules, and
ortho-to-para ratios for vibrationally excited molecules,
have been observed in many PDRs.
Such sources include the reflection nebulae NGC 2023 and NGC 7023
(Gatley et al. 1987; Hasegawa et al. 1987; Martini, Sellgren \& Hora 1997), 
the planetary nebulae Hubble 12, NGC 7027 and 
BD+30$^\circ$3639 (Tanaka et al. 1989;
Ramsay et al. 1993; Hora \& Latter 1996;
Luhman \& Rieke 1996;
Shupe et al. 1998), PDRs in the star-forming
region M17 (Chrysostomou et al. 1993 et al.),
the bipolar outflow DR21 (Fernandes, Brand \& Burton 1997),
and the starburst galaxy NGC 253 (Harrison et al. 1998).
Typically, in regions where FUV-pumping dominates the
molecular excitation the vibrationally excited ortho-to-para ratios 
lie in the range 1.5-2.2.

Contrary to what has been implicitly assumed in much of
the previous literature on this subject, our results show that the actual
abundance ratio of ortho and para molecules in PDRs is
not accurately represented by the ortho-to-para
ratio determined from observations of the vibrational emission lines.
Previous discussions
have erroneously assumed that the ortho-to-para ratios inferred from
such observations
must be explained by actual ortho
to para-H$_2$ ratios smaller than 3, and have invoked
a variety of processes that might lead to such ratios,
including the formation of $\rm H_2$
with an initial ortho-to-para ratio less than 3,
rapid ortho-para conversion in low-temperature gas,
or warm but time-dependent PDRs in
which the ortho-to-para ratio has not had sufficient
time to reach its high temperature
LTE value.  Our results indicate that such
explanations are unnecessary: most observed
values of the vibrationally
excited ortho-to-para ratio are entirely consistent with the
assumption that the true ortho-to-para abundance ratio is in fact 3.

Recent {\it Infrared Space Observatory (ISO)} observations
of the PDR in the star-forming molecular cloud
S140 corroborate our central result. With
a broad spectral coverage that is not affected
by atmospheric absorption, {\it ISO} has allowed
both vibrational emissions {\it and} pure
rotational mid-IR transitions of H$_2$ to be measured from S140
(Timmermann et al.\ 1996).  These observations
show that the relative intensities of the
vibrational emissions in S140 are consistent with excitation via
FUV-pumping, but that the pure rotational emissions 
are collisionally excited in warm gas at a temperature $\sim 500$ K.
Most importantly, the ortho-to-para ratio for vibrationally
excited molecules is observed to be significantly less
than 3, whereas the pure rotational emissions reveal that
the true ortho-to-para H$_2$ abundance ratio is equal to an LTE value of 3.

In Fig. 5 we present a model fit to the S140 data which illustrates
this key observational fact.  In our model we assume that 
$n=10^4$ cm$^{-3}$ and $\chi=500$, and that in S140 the cloud surface is
inclined to the line-of-sight by an 
angle $\theta$ with ${\rm cos}\,\theta=0.1$.
As discussed by 
Timmermann et al. (1996), these are the appropriate values for the
gas density, incident FUV field intensity, and cloud inclination
for the S140 PDR.
For these parameters our model provides an excellent fit
to the observed column densities of vibrationally excited
molecules.
In our model we also assume that the gas temperature varies as
\footnote{Other functional forms for the thermal profile may be assumed.
For example, in the model presented by Timmermann et al. (1996)
the gas temperature is assumed to vary as $T= 20K + 980K/(1+\tau^2)$,
where $\tau$ is the FUV dust continuum optical depth.}
$$
T = {{\rm 500 \ K}\over 1+9(2n({\rm H}_2)/n)^4}
 \ \ \ . \eqno(19)
$$
For this thermal profile the gas temperature
is equal to 500 K in the outer atomic layer, and falls rapidly to 50 K as the
gas becomes molecular.
In this model we adopt the same H-H$_2$ and H$^+$-H$_2$ rate coefficients
and proton abundance as in the third model presented in \S 3.
Our S140 model
provides a very good fit to the observed column densities of 
rotationally excited molecules in the $v=0$ state.
This suggests that 
in S140 the warm gas traced by the rotational emissions
is confined mainly to cloud layers where
the hydrogen is atomic or partially dissociated.

In our model the ortho-to-para column density 
ratio for vibrationally excited
molecules converges to a value of 2. Fig. 5 shows that this
ratio is formed in outer cloud layers where the gas temperature
$\gtrsim 200$ K, and where the true ortho-to-para
abundance ratio is close to the LTE value of 3. Fig. 5 shows that
at cloud depths
$A_V\gtrsim 0.3$ the ortho-to-para abundance ratio falls
below 3 as the gas becomes cold. However, at these large
cloud depths the FUV-pumping rates become vanishingly small, and
the total column densities of vibrationally excited molecules are unaffected.
The excitation diagram in Fig. 5
(see also Fig. 3 of Timmermann et al. 1996) displays the
observed values of $(N_{vj}/g)$ in S140 together with our
model fit. The excitation diagram again shows 
that the ortho-to-para ratio $\sim 2$ in vibrationally excited states
is significantly less than the true LTE value of 3 indicated by the
pure rotational states. Our S140 model is similar to the models
displayed in Figs. 2 and 4, and the behavior of the total
and vibrationally excited
ortho-to-para ratios in S140 is consistent with equation (13).

The S140 observations provide strong support for our conclusion
that the ortho-to-para ratios of $\sim 1.7$ typically inferred
from observations of FUV-pumped H$_2$ emissions imply
that the true ortho-to-para
abundance ratios in the emitting regions are equal to LTE values of 3.
An LTE ortho-to-para abundance ratio less than 3 would only be
indicated by a vibrationally excited ortho-to-para ratio
significantly less than $\sqrt{3}\sim 1.7$

In shock-heated regions or in very hot PDRs, where 
collisional excitation is dominant, $\rm H_2$ vibrational emissions 
are expected to reflect the true ortho-to-para ratio.
In fact, observations of $\rm H_2$ vibrational
emissions in shocked regions associated with
protostellar outflows and Herbig Haro objects
do indeed reveal vibrationally excited ortho-to-para
ratios close to 3 (Smith et al.\ 1997). 
The effects of collisional excitation to $v=1$ within a
very hot PDR are elegantly demonstrated by the recent observations of 
of the planetary nebula BD $+30^\circ 3639$ by Shupe et al. (1998).
Spatially-resolved line ratios in this source 
indicate a decrease in the $v=2-1$ to $v=1-0$ line ratio and
an increase in the ortho-to-para ratio for $v=1$ (from a
value $\sim 1.7$ to a value
$\sim 3$) with decreasing distance from the
central star.  Both effects may be interpreted as resulting
from collisional excitation to $v=1$ in the warmest parts of the
nebula lying closest to the star, and FUV-pumping of $v=1$ 
molecules in the outer cooler portions of the nebula.
Our analysis shows that the true ortho-to-para ratio
is likely very close to 3 at {\it all} observed locations in this source.

\section{Discussion}

Several theoretical discussions of the ortho-to-para ratio in PDRs have
been presented in the literature.
Takayanagi, Sakimoto \& Onda (1987) argued that the ortho-to-para
ratio of $\sim 1.7$ inferred by Hasegawa et al. (1987) from observations of
fluorescent H$_2$ emission in 
NGC 2023 implies that either the gas temperature $T_{gas}$ or formation
temperature $T_{form}$ must be very low ($\lesssim 70$ K).
However, as pointed out already by Black \& van Dishoeck (1987), 
Takayanagi et al. (1987) assumed that the FUV absorption lines
are optically thin, and neglected the critically important
effects of molecular self-shielding.  The computations of
Takayanagi et al. (1987) are therefore not relevant to the
interpretation of the ortho-to-para ratios in realistic PDRs.  
Black \& van Dishoeck (1987) also
noted that in their own models (which do
include optical depth effects in the FUV absorption lines)
low vibrational ortho-to-para ratios are obtained even when
the formation temperature is large and $\alpha(T_{form})=3$,
and suggested that their results may be due in part to
different depth dependences of the ortho and para pumping rates.

More recently, Draine \& Bertoldi (1996) presented detailed models of 
$\rm H_2$ excitation and emission in PDRs, and presented explicit computations of 
the ortho-to-para ratios for vibrationally excited ($v=1$) molecules for a 
wide range of cloud densities, FUV field intensities, and gas temperatures. 
Ortho-to-para ratios were obtained for the vibrational-excited states
of H$_2$ for two sequences of models -- one with the ratio $\chi/n=0.1$ 
cm$^3$ and another for $\chi/n=0.01$ cm$^3$ -- and for each sequence results 
were presented for $n$ ranging from 10$^2$ to 10$^6$ cm$^{-3}$, 
and maximum gas temperatures ranging from 200 to 800~K.  

The numerical results of Draine \& Bertoldi, as summarized in Fig. 17 of their
paper, are fully consistent with our analysis, although the authors did not
explicitly discuss the crucial distinction between the ortho-to-para ratio 
in vibrationally-excited states and the true ortho-to-para abundance ratio:
that distinction that seems to have been misunderstood in much of the 
previous and subsequent literature.

Draine \& Bertoldi found that for $n\lesssim 10^4$ cm$^{-3}$ and 
$\chi \simlt 100$, the vibrationally excited ortho-to-para ratio is always 
equal to a non-LTE value of 
$2\pm 0.2$, significantly less than 3, while at higher densities and FUV 
fields the ratio approaches and even exceeds the LTE value of 3.
Our analysis shows that the behavior exhibited by the Draine \& Bertoldi 
(1996) results at low $n$ and $\chi$ is almost certainly due to the
preferential reduction of the pumping rates for ortho-H$_2$
which as we have discussed is expected for radiative fluorescent H$_2$ emission.
\footnote{Draine \& Bertoldi (1996) did not present the values of 
the ortho-to-para ratios for $v=0$ molecules in their models, but our 
analysis suggests that these must be close to 3.}  The behavior for
high $n$ and $\chi$, by contrast, is likely due to the combined effects of
collisional excitation of $v=1$ molecules in hot gas,
as well as 
of preferential self-shielding of vibrationally excited ortho molecules 
as discussed in \S 2
and suggested also by Draine \& Bertoldi. 

In conclusion, in our paper we have addressed a long standing confusion
between the ortho-to-para ratio of H$_2$ molecules
in FUV-pumped vibrationally excited
states, and the actual ortho-to-para H$_2$ abundance ratio.
We have argued that because the process of FUV-pumping in PDRs occurs via
optically thick absorption lines, a typically observed ortho-to-para ratio
of $\sim 1.7$ for FUV-pumped molecules
is consistent with, and implies, an actual LTE ortho-to-para abundance
ratio of 3, and is a signature of
warm gas. Our conclusions are strongly supported
by recent {\it ISO} observations of the PDR in S140.

A.S. thanks the Radio Astronomy Laboratory at
U.C. Berkeley, the Center for Star Formation Studies
consortium, and the German-Israeli Foundation (grant I-551-186.07/97)
for support. D.A.N. gratefully acknowledges the support
of NASA grant NAG5-3316.

\vfill\eject
\centerline{\bf References}
\vskip 0.1 true in 
{\hoffset 20pt
\parindent = -20pt

Allison, A.C., \& Dalgarno, A.\ 1970, Atomic Data Tables, 1, 289

Black, J.H., \& Dalgarno, A.\ 1976, ApJ, 203, 192

Black, J.H., \& van Dishoeck, E.F., ApJ, 1987, 322, 412

Boothroyd, A.I., Keough, W.J., Martin, P.G., \& Peterson, M.R.\
1991, J.Chem.Phys., 95, 4343

Burton, M.G., Hollenbach, D.J., \& Tielens, A.G.G.M.\ 1990, ApJ, 365, 620

Burton, M.G., Hollenbach, D.J., \& Tielens, A.G.G.M.\ 1992, ApJ, 399, 563 

Chrysostomou, A., Brand, P.~W.~J.~L., Burton, M.~D.,
\& Moorhouse, A. 1993, MNRAS, 265, 329

Dalgarno, A., Black, J., \& Weisheit, J.C.\ 1973, Ap.Lett., 14, 77

Draine, B.T.\ 1978, ApJS, 36, 595

Draine, B.T., \& Bertoldi, F.\ 1996, ApJ, 468, 269

Federman, S.R., Glassgold, A.E., \& Kwan, J.\ 1979, ApJ, 227, 466

Fernandes, A.J.L., Brand, P.W.J.L., \& Burton, M.G.\ 1997, MNRAS, 290, 216

Flower, D.R., \& Watt, G.D.,\ 1984, MNRAS 209, 25

Gatley, I., et al.~1987, ApJ, 318, L73

Gerlich, D.\ 1990, J.Chem.Phys., 92, 2377

Harrison, A., Puxley, P., Russell, A., \& Brand, P.\ 1998, MNRAS, 297, 624

Hasegawa, T., Gatley, I., Garden, R.P, Brand, P.W.J.L., 
Ohishi, M., Hayashi, M., \& Kaifu, N.\ 1987, ApJ, 318, L77

Hollenbach, D.J., \& Tielens, A.G.G.M.\ 1997, ARAA, 35, 179

Hollenbach, D.J., \& Tielens, A.G.G.M.\ 1998, Rev.Mod.Phys. in press

Hora, J.L., \& Latter, W.B. 1996, ApJ, 461, 288 

Lepp, S., Tin\`e, S., \& Dalgarno, A.\ 1998, in prep.

Luhman, K.L., \& Rieke, G.H.\ 1996, ApJ, 461, 298

Luhman, M.L., Jaffe, D.T., Sternberg, A., Herrmann, F., \&
Poglitsch, A.\ 1997, ApJ, 482, 298

Maloney, P., Hollenbach, D.J., \& Tielens, A.G.G.M.\ 1996, ApJ, 466, 561

Martin, P.G., \& Mandy, M.E.\ 1993, ApJS, 86, 199

Martini, P., Sellgren, K., \& Hora, J.L.\ 1997, ApJ, 484, 296

Neufeld, D.A., \& Spaans, M.\ 1996, ApJ, 473, 894

Neufeld, D.A., Melnick, G.J., \& Harwit, M.\ 1998, ApJ, 506, L75


Ramsay, S.K., Chrysostomou, A., Geballe, T.R., Brand, P.W.J.L,
\& Mountain, M.\ 1993, MNRAS, 263, 695

Shupe, D.L., Larkin, J.E., Knop, R.A., Armus, L., Matthews, K., \& 
Soifer, B.T.\ 1998, ApJ, 498, 267

Smith, M.D., Davis, C.J., \& Lioure, A.\ 1997, A\&A, 327, 1206 

Sternberg, A.\ 1988, ApJ, 332, 400

Sternberg, A.\ 1998, in The Molecular Astrophysics of Stars and Galaxies,
ed. T.W. Hartquist and D.A. Williams (Oxford University Press)

Sternberg, A., \& Dalgarno, A.\ 1989, ApJ, 338, 197 

Sternberg, A., \& Dalgarno, A.\ 1995, ApJS, 99, 565

Stephens, T.L., \& Dalgarno, A.\ 1972, J. Quant. Spectrosc. Radiat. Transfer,
12, 569

Takayanagi, K., Sakimoto, K., \& Onda, K.,\ 1987, ApJ, 318, L81

Tanaka, M., Hasegawa, T., Hayashi, S.S., Brand, P.W.J.L., \& Gatley, I.\
1989, ApJ, 336, 207	

Tielens, A.G.G.M., \& Hollenbach, D.J. 1985, ApJ, 291, 722

Timmermann, R., Bertoldi, F., Wright, C.M., Drapatz, S., 
Draine, B.T., Haser, L., \& Sternberg, A.\ 1996, 
A\&A, 315, L281

Tin\'e, S., Lepp, S., Gredel, R., \& Dalgarno, A.\ 1997, ApJ, 481, 282.

Trulhar, D.G., \& Horowitz, C.J.\ 1978, J.Chem.Phys., 68, 2466

van Dishoeck, E.F., and Black, J.H.\ 1987, ApJ, 322, 412

Wolniewicz, L., Simbotin, I, \& Dalgarno, A.,\ 1998, ApJS, 115, 293

}
\vfill\eject
  
\centerline{\bf Figure Captions}
\vskip 0.1 true in 
{\hoffset -20pt
\parindent -20pt

Fig. 1 -- The ratio of ortho- to para-H$_2$ in local thermodynamic
equilibrium, $\alpha(T_{gas})$, as a function of the gas temperature,
$T_{gas}$.

Fig. 2 -- (a) Top left panel: The atomic hydrogen fraction $n(H)/n$,
and the column density of vibrationally excited H$_2$
relative to the total column of vibrationally excited H$_2$ formed in
the PDR,  $N^*/N^*_{tot}$, as functions of visual extinction, A$_V$.
The total hydrogen gas density $n=10^4$ cm$^{-3}$,
the incident FUV radiation intensity is $\chi=2\times 10^3$,
and the gas temperature $T_{gas}=500$ K.
In this model the rotational levels in the v=0 level are
assumed to be in LTE everywhere in the cloud.

(b) Bottom left panel: The total and vibrationally excited 
ortho-to-para ratios.
The dashed curves are the local ortho-to-para ratios. 
The solid curves are the integrated column ortho-to-para ratios.

(c) Right panels: An excitation diagram showing $log(N/g)$ vs. $E$,
where $N$ are the total model column densities in each of the 
rotational-vibrational levels, $g$ is the statistical weight,
and $E$ is the energy (in K) of the level.

Fig. 3 -- In this model ortho-para interconversions for the
$v=0$ molecules is suppressed.

Fig. 4 -- In this model realistic ortho-para conversions via
$\rm H^+-H_2$ and $\rm H-H_2$ collisions are included in the computations
(see text). 

Fig. 5 -- A model for the PDR in S140, with $n=10^4$ cm$^{-3}$, $\chi=500$,
and inclination $cos\theta=0.1$.

(a) Top left panel: 
The atomic hydrogen fraction $n(H)/n$,
and the column density of vibrationally excited H$_2$, $N^*/N^*_{tot}$,
and the temperature profile $T/T_{max}$ (see text)
as functions of visual extinction, A$_V$.

(b) Bottom left panel: The total and vibrationally excited
ortho-to-para ratios.
The dashed curves are the local ortho-to-para ratios.
The solid curves are the integrated column ortho-to-para ratios.

(c) Right panel: An excitation diagram showing the observed
and model values of $N/g$.  The observations are indicated
by the errorbars assuming $\pm$30\% ISO-SWS flux calibration
uncertainties (see Timmermann et al. 1996). The square symbols
indicate the model values for pure rotational states, and the 
triangles indicate vibrationally excited states.  The
dashed line indicates LTE gas at 500 K.
} 

\clearpage
\begin{figure}[htpb]
\plotone{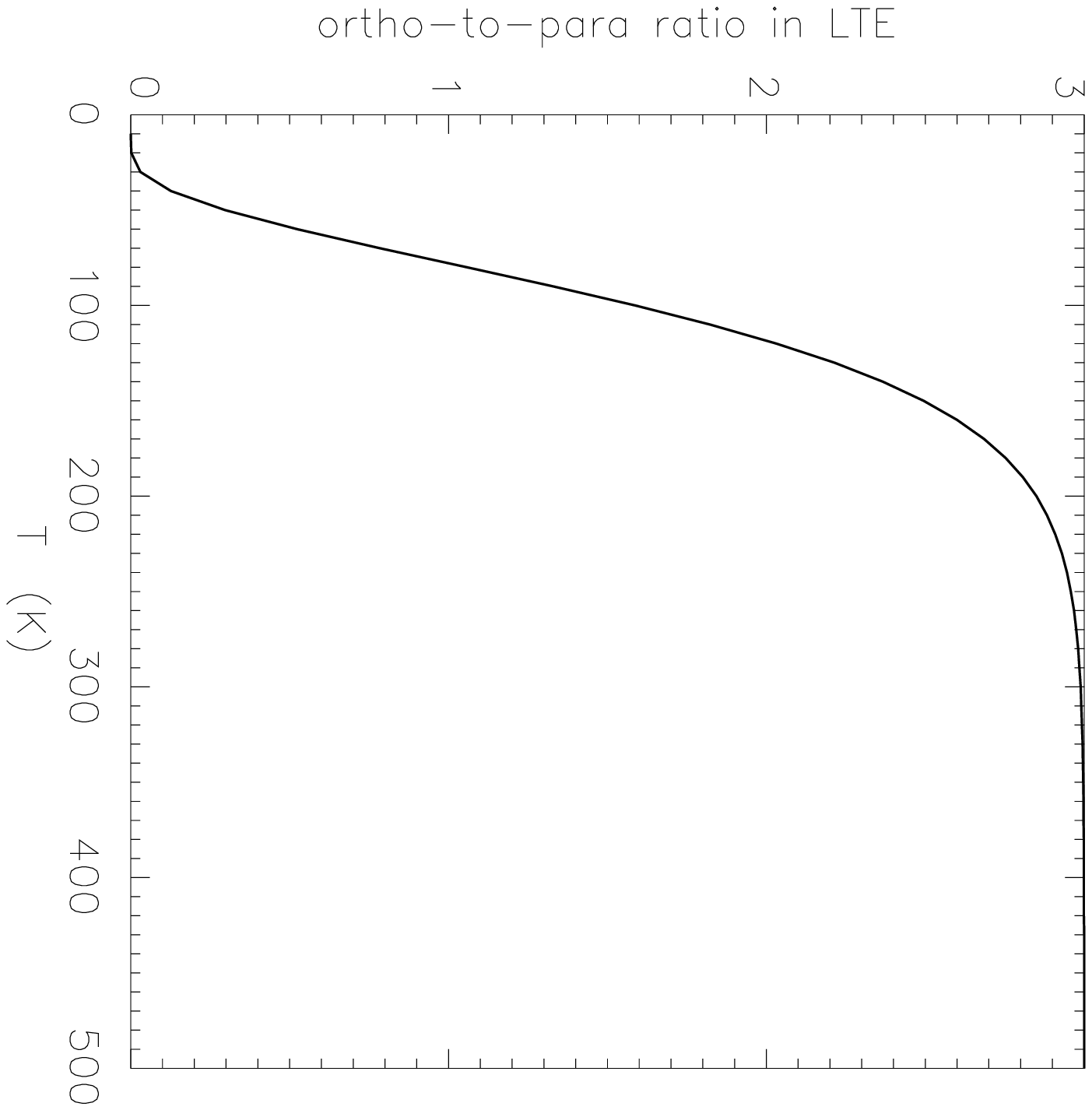}
\end{figure}

\clearpage
\begin{figure}[htpb]
\plotone{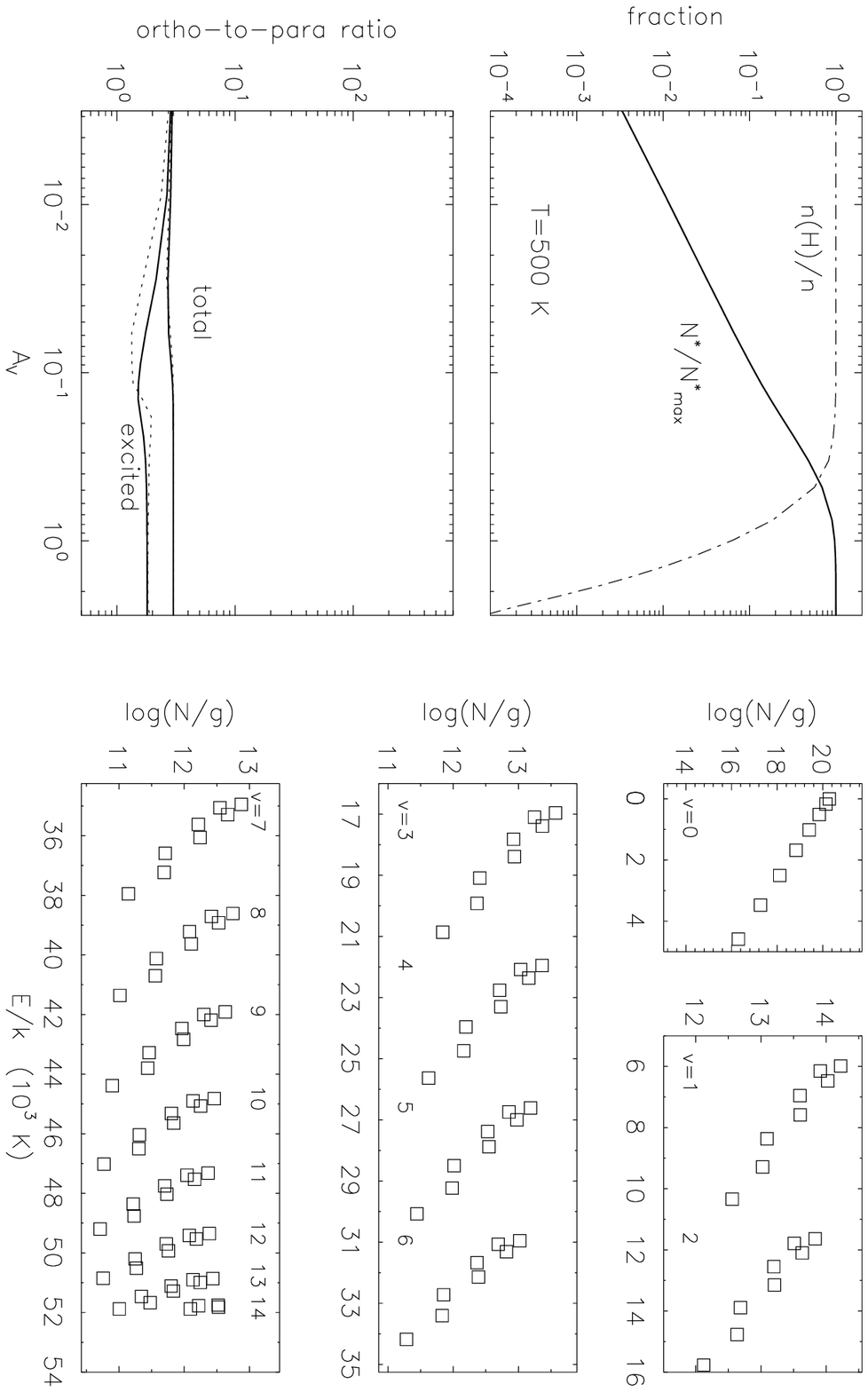}
\end{figure}

\clearpage
\begin{figure}[htpb]
\plotone{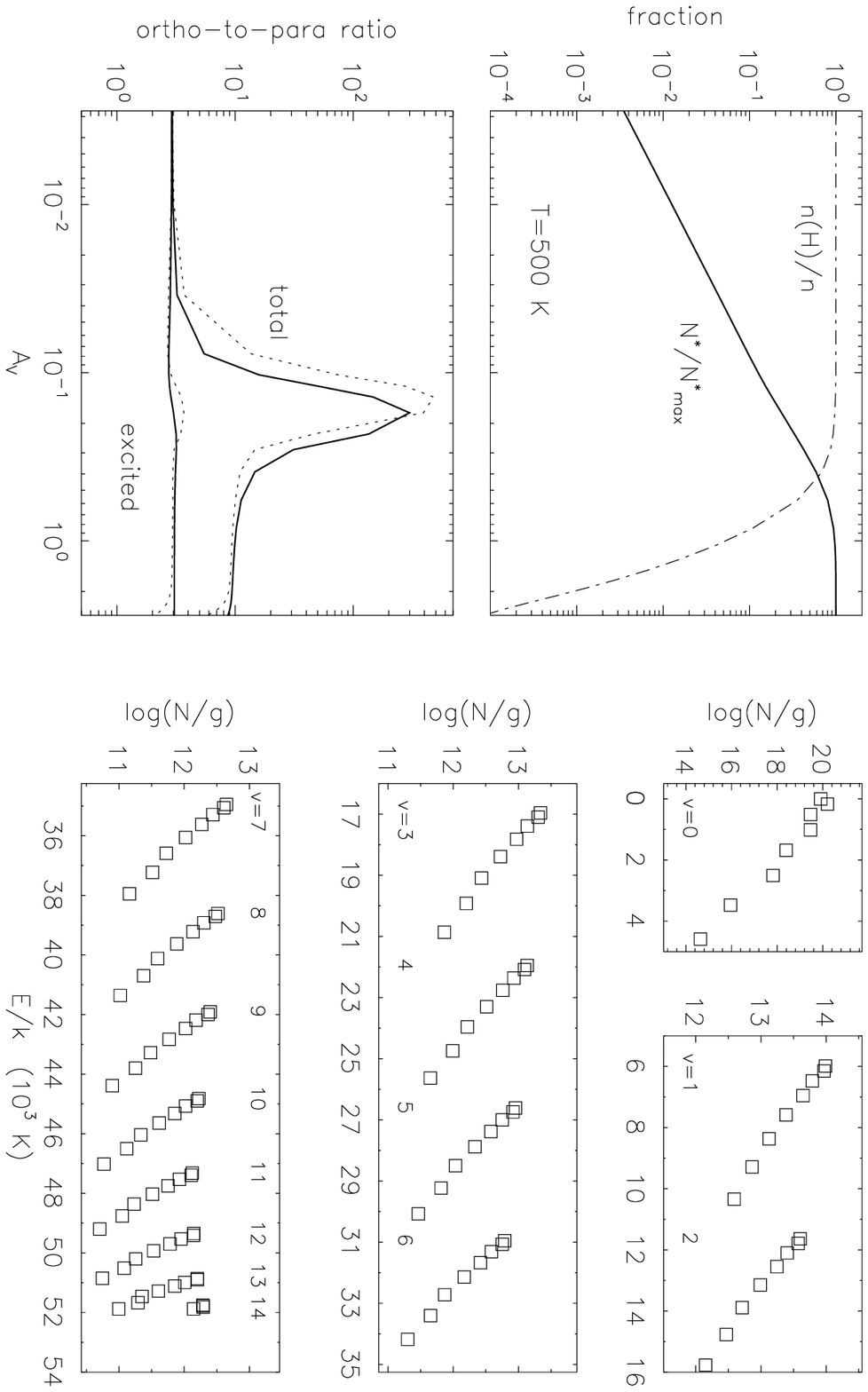}
\end{figure}

\clearpage
\begin{figure}[htpb]
\plotone{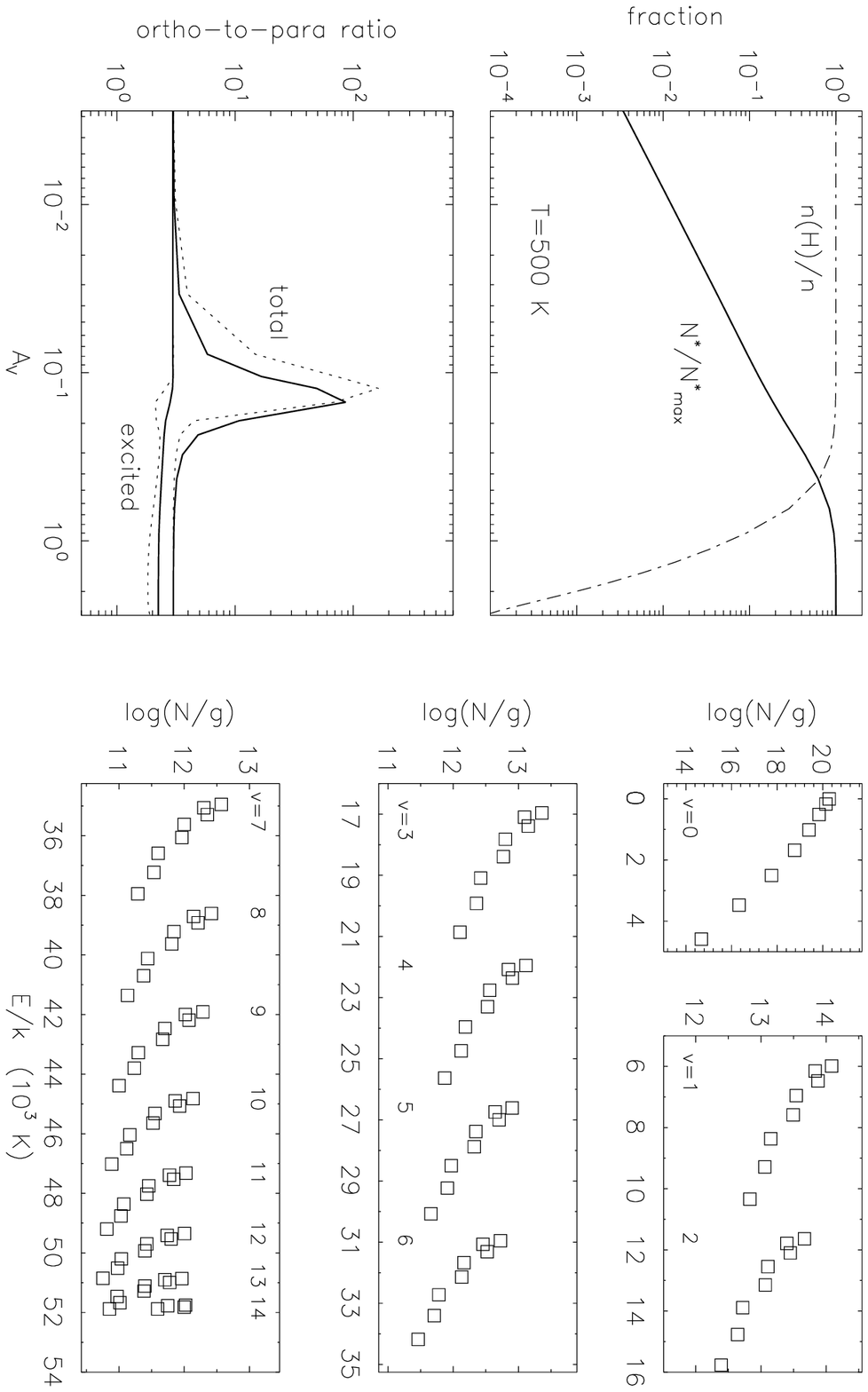}
\end{figure}

\clearpage
\begin{figure}[htpb]
\plotone{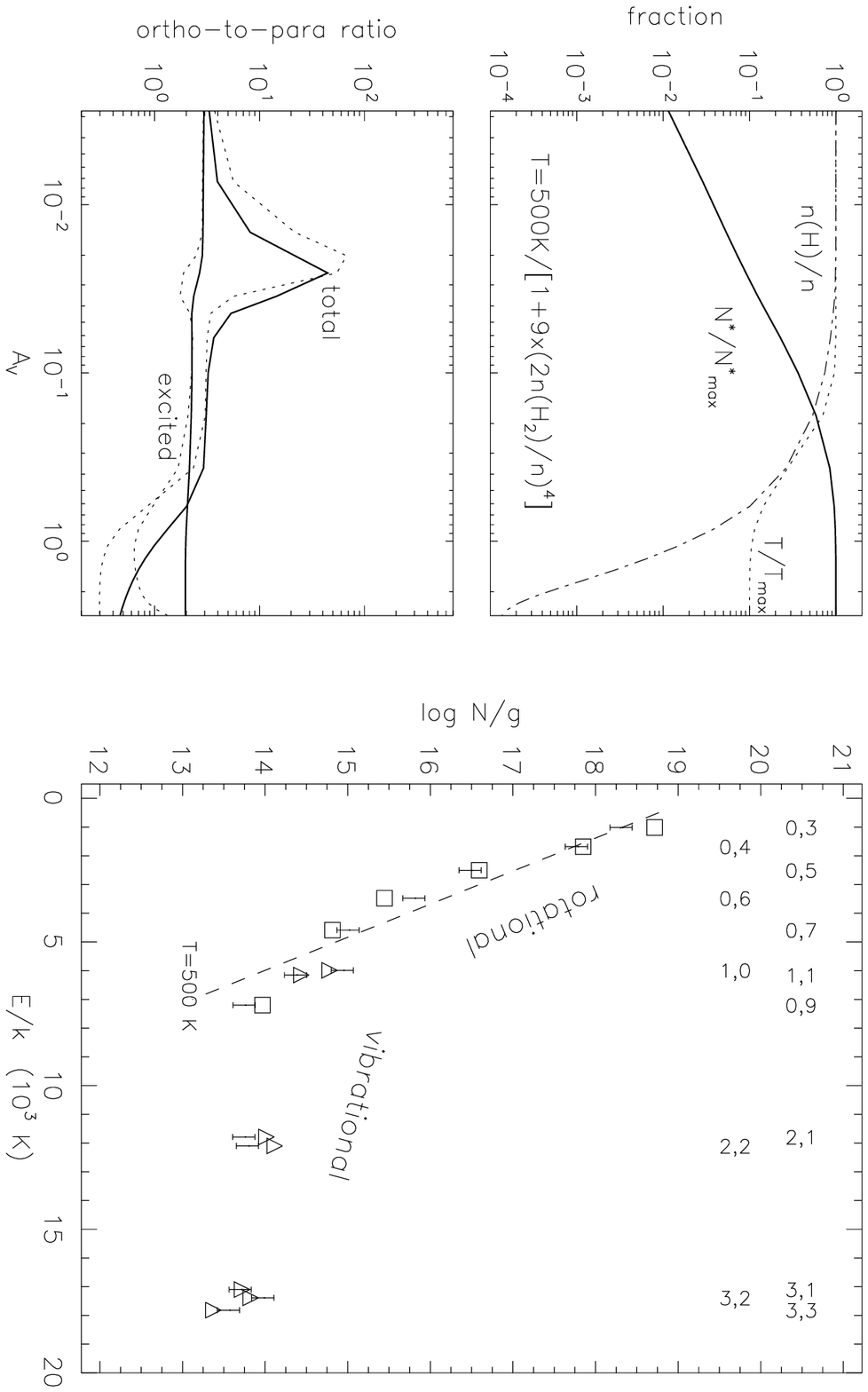}
\end{figure}

\end{document}